\documentclass[reprint,aps,prb,superscriptaddress,a4paper,nofootinbib,showpacs,showkeys,floatfix]{revtex4-1}

\setcitestyle{numbers,square}

\usepackage{graphicx,epsfig,epstopdf}
\usepackage{amssymb,amsmath,bm,dcolumn,braket,url,color}
\usepackage[normalem]{ulem}

\newcommand{\new}[1]{{ #1}}

\newcommand{\mm}[1]{\mbox{$#1$}}

\newcommand{\ms}{\mbox{$\mu_{\mathrm{spin}}$}}
\newcommand{\mo}{\mbox{$\mu_{\mathrm{orb}}$}}

\newcommand{\emca}{\mbox{$E_{\text{MCA}}$}}

\newcommand{\lfe}{\mbox{$\lambda_{\text{Fe}}$}}
\newcommand{\lpt}{\mbox{$\lambda_{\text{Pt}}$}}

\newcommand{\mca}{magnetocrystalline anisotropy}

\newcommand{\sopt}{second order perturbation theory}
\newcommand{\xcf}{exchange-correlation functional}

\newcommand{\kkr}{{\sc sprkkr}}
\newcommand{\wien}{{\sc wien}2k}

\newcommand{\ea}{{\it et al.}}

\begin{document}

\title{Magnetocrystalline anisotropy of FePt: a detailed view}

\author{Saleem Ayaz \surname{Khan}} \affiliation{New Technologies
  Research Centre, University of West Bohemia, Univerzitn\'{\i} 2732,
  306 14 Pilsen, Czech Republic}

\author{Peter \surname{Blaha}} \affiliation{Institute of Materials
  Chemistry, TU Vienna, Getreidemarkt 9, A-1060 Vienna, Austria}

\author{Hubert \surname{Ebert}} \affiliation{Universit\"{a}t
  M\"{u}nchen, Department Chemie, Butenandtstr.~5-13,
  D-81377~M\"{u}nchen, Germany}

\author{Jan \surname{Min\'{a}r}} \affiliation{New Technologies
  Research Centre, University of West Bohemia, Univerzitn\'{\i} 2732,
  306 14 Pilsen, Czech Republic} \affiliation{Universit\"{a}t
  M\"{u}nchen, Department Chemie, Butenandtstr.~5-13,
  D-81377~M\"{u}nchen, Germany}

\author{Ond\v{r}ej \surname{\v{S}ipr}}  \affiliation{New Technologies
  Research Centre, University of West Bohemia, Univerzitn\'{\i} 2732,
  306 14 Pilsen, Czech Republic}  \affiliation{Institute of Physics
  ASCR v.~v.~i., Cukrovarnick\'{a}~10, CZ-162~53~Prague, Czech
  Republic }

%%%%%%%%%%%%%%%%%%%%%%%%%%%%%%%%%%%%%%%%%%%%%%%%%%%%%%%%%%%%%%%%%%

\date{\today}

\begin{abstract}
To get a reliable ab-initio value for the magneto-crystalline
anisotropy (MCA) energy of FePt, we employ the full-potential
linearized augmented plane wave (FLAPW) method and the full-potential
Korringa-Kohn-Rostoker (KKR) Green function method. The MCA energies
calculated by both methods are in a good agreement with each other.
As the calculated MCA energy significantly differs from experiment, it
is clear that many-body effects beyond the local density approximation
are
\new{essential.  It is not really important whether relativistic
  effects for FePt are accounted for by solving the full Dirac
  equation or whether the spin-orbit coupling (SOC) is treated as a
  correction to the scalar-relativistic Hamiltonian.  }
From the analysis of the dependence of the MCA energy on the
magnetization angle and on the SOC strength it follows that the main
mechanism of MCA in FePt can be described within \sopt.  However, a
\new{distinct } contribution not accountable for by \sopt\ is present
as well.
\end{abstract}


\maketitle

%%%%%%%%%%%%%%%%%%%%%%%%%%%%%%%%%%%%%%%%%%%%%%%%%%%%

\section{Introduction}   \label{sec-intro}

The various ab-initio electronic structure codes use different
approaches to solve the Schr\"{o}dinger equation for a solid.  Usually
different codes and/or methods yield results that are similar but show
sometimes important differences in the details.  These details start
to matter if one aims at high-precision calculations with predictive
power. Therefore an effort has lately intensified to standardize
ab-initio calculations and to find the conditions that have to be met
so that reliable ``true'' quantitative values are obtained.  So far
the attention has been paid mostly to total energies, equilibrium
lattice parameters and bulk moduli
\cite{ASH+99,Kurt+14,deltacodes,LBB+15}. We want to extend this effort
to another numerically sensitive area, namely, to the
magneto-crystalline anisotropy (MCA).

The MCA is manifested by the fact that the energy of a magnetically
ordered material depends on the direction of the magnetization
$\bm{M}$ with respect to the crystal lattice.  It is an interesting
phenomenon both for fundamental and technological reasons, as the MCA
is important among others for the design of magnetic recording media.
\new{Theoretical research on MCA proceeds in two directions. First, one tries
to understand the mechanism behind the MCA in simple intuitive terms, so that one
would have guidance in search for materials with a high MCA
energy \cite{vdL+98,KW+14,SMP+16}. Second, one tries to find which
computational procedures have to be employed so that one can make
quantitative predictions on the MCA energy
\cite{JSW+03,Burkert+05,BH+09}. }

Getting an accurate value of the MCA energy $E_{\text{MCA}}$ is quite
difficult as one has to, at least in principle, subtract two very
large numbers (total energies for two orientations of magnetization)
to get a very small number, namely, $E_{\text{MCA}}$.  Several
conditions for getting accurate well-converged results were explored
in the past.  In particular, the importance of a sufficiently dense
mesh in the Brillouin zone (BZ) for the $\bm{k}$-space integration was
recognized \cite{GR+86,RKF+01,Singh+03}. When dealing with supported
systems such as adatoms or monolayers, the semi-infinite substrate has
to be properly accounted for \cite{SBM+10,SBE+14}. Despite all the
efforts, getting accurate and reliable predictions of the MCA energy
is still a problem.  
\new{Numerical uncertainties severely restrict the practical
  usefulness of calculations of the MCA. They hinder our understanding
  of the underlying physics as well, because lack of reliable
  numerical values means that it is not really possible to determine
  which physical approximations and models are acceptable and which
  are not. }

In this work we focus on MCA of bulk FePt.  This compound has the
largest MCA energy of all bulk materials formed by transition metals
and its crystal structure is quite simple, so it is a good candidate
for a reliable calculation. At the same time, the presence of Pt --- a
heavy element --- suggests that relativistic effects should be
significant, offering thus an interesting possibility to check how
different methods of dealing with relativistic effects, in particular
with the spin-orbit coupling, influence the results.  
\new{Besides, a deeper understanding of the MCA of FePt is important
  regarding current search for suitable rare-earth-free magnetic
  materials.  Transition metals are natural candidates in this respect
  and attracted a lot of attention recently
  \cite{Jiangfeng+11,AGS+11}. }

Previous theoretical studies on FePt based on the local density
approximation (LDA) give a large spread of the results --- from
1.8~meV to 4.3~meV
\cite{Daalderop+91,Sakuma+94,Solovyev+95,Oppeneer+98,Galanakis+00,RKF+01,Shick+03,Ostanin+03,Burkert+05}.
If one restricts to full potential methods only, one still gets a
relatively large difference between various studies: \emca\ of FePt
was determined as 2.7~meV by FP-LMTO calculation of Ravindran
\ea\ \cite{RKF+01} and FLAPW calculation of Shick and Mryasov
\cite{Shick+03}, 3.1~meV by plane-waves calculation of Kosugi
\ea\ \cite{KMI+14}, and 3.9~meV by FP-LMTO calculation of Galanakis
\ea\ \cite{Galanakis+00}. The differences between various LDA
calculations are comparable to the differences between LDA results and
the experimental value of 1.3 meV \cite{Ivanov+73}. Even though part
of the spread of the LDA results can be attributed to the use of
different LDA exchange-correlation functionals, the differences are
still too large to be acceptable.  Besides, they occur also for
studies which use the same \xcf\ (e.g., both Ravindran
\ea\ \cite{RKF+01} and Galanakis \ea\ \cite{Galanakis+00} use von
Barth and Hedin functional \cite{Barth+72}). This suggests that the
accuracy of ab-initio MCA energy calculations may not \new{even } be
sufficient to answer \new{the fundamental } question whether the LDA
itself is able to reproduce the experimental MCA energy of FePt or
not.

Deciding which method gives better MCA results than the other is quite
difficult, among others because different computational approaches
used by different codes are intertwined with different ways of
implementing relativistic effects.  Recall that as the MCA is
intimately related to the spin orbit coupling (SOC), the way the
relativity is included can be an important factor.
\new{To verify that a calculated MCA energy really represents the true LDA
  value, one has to use two different methods and make sure that the
  calculations are properly converged.  }

The aim of our work is to perform a robust and reliable LDA
calculation of the MCA energy of FePt 
\new{to find out whether treating relativistic effects via an
  explicit SOC Hamiltonian is sufficient for MCA calculations,}
whether the MCA of FePt can be described within the LDA, 
\new{and what is the accuracy of current MCA energy calculations in
  general.}

The first computational method we employ is the well-established and
recognized full potential linearized augmented plane wave (FLAPW)
method as implemented in the {\sc wien2}k code \cite{Blaha+01}.  This
method was used as a reference in the recent study of the accuracy of
total energies and related quantities
\cite{Kurt+14,deltacodes,LBB+15}.
\new{Relativistic effects are treated approximately in \wien,
  by introducing a separate SOC-related term to the Hamiltonian. }
As the second method we opted for a fully relativistic 
\new{full  potential } 
multiple scattering KKR (Korringa-Kohn-Rostoker) Green
function formalism as implemented in the {\sc sprkkr} code
\cite{Ebert+11,sprkkr+12}.

Many aspects of the MCA of FePt were theoretically investigated in the
past already.  Daalderop \ea\ \cite{Daalderop+91} and Ravindran
\ea\ \cite{RKF+01} studied the influence of the band-filling on
$E_{\text{MCA}}$ of FePt. Many groups studied the influence of the
temperature on the MCA of FePt \cite{Mryasov+05,Sta+04}. The
dependence of the Curie temperature on the FePt grain size was
investigated via model Hamiltonian calculations
\cite{Hovorka+12}. Burkert \ea\ \cite{Burkert+05}, Lukashev
\ea\ \cite{Lukashev+12} and Kosugi \ea\ \cite{KMI+14} studied how
$E_{\text{MCA}}$ depends on the strain (i.e., the $c$/$a$
ratio). Cuadrado \ea\ \cite{Cuadrado+14} gradually substituted the Fe
atom by Cr, Mn, Co, Ni, or Cu to find that the MCA energy of
Fe$_{1-y}$X$_{y}$Pt alloys can be tuned by adjusting the content of
the substituting element.

To facilitate the understanding of the MCA of FePt and related systems
further, we focus on some aspects that have not been paid attention so
far.
\new{We show that if \wien\ and \kkr\ calculations are converged, they
  yield comparable values for the MCA energy.  Dealing with relativity
  by introducing an additional SOC-related term to the Hamiltonian is
  thereby justified. The theoretical MCA energy of FePt (3.0~meV) is
  significantly larger than the experimental value (1.3 meV), implying
  conclusively that the LDA cannot properly describe the MCA of FePt.
  We also analyze how the total energy varies with the magnetization
  angle and how MCA energy scales with the SOC strength. Based on this
  we conclude that even though the MCA of FePt is dominated by a
  \sopt\ mechanism, there is a small but distinct contribution
  originating from the Pt sites which is not accountable for by
  \sopt. }

%%%%%%%%%%%%%%%%%%%%%%%%%%%%%%%%%%%%%%%%%%%%%%%%%%%%

\section{Computational details}   \label{sec-comput}

\subsection{Technical details}   \label{sec-dets}

\begin{figure}
\includegraphics [width=0.6\linewidth]{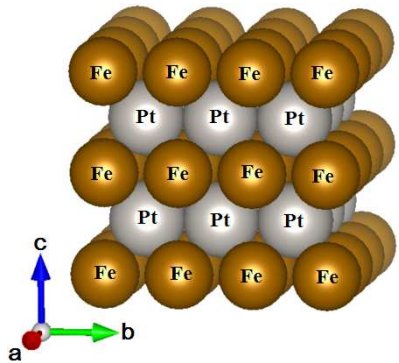}
\caption{Crystal structure of bulk L1$_{0}$ FePt \label{FePt-cell}}
\end{figure}

We studied bulk FePt with the L1$_{0}$ layered structure.  The lattice
parameters of a $tP2$ unit cell are $a$=2.722~\AA\ and $c$=3.714~\AA.
Fe atoms and Pt atoms are at the (0.0,0.0,0.0) and (0.5,0.5,0.5)
crystallographic positions, respectively, resulting in a compound with
alternating Fe and Pt atomic layers stacked along the $c$ axis (see
Fig.~\ref{FePt-cell}).

We used two different computational methods, namely, the FLAPW method
as implemented in the {\sc wien2}k code \cite{Blaha+01} and the
multiple scattering KKR Green function method as implemented in the
{\sc sprkkr} code \cite{Ebert+11,sprkkr+12}. Our calculations are based
on the LDA.  The values presented in the section Results
(Sec.~\ref{sec-res}) were obtained using the Vosko, Wilk and Nusair
(VWN) \xcf\ \cite{VWN+80}. Use of different LDA functionals leads to
small but identifiable changes in \emca, as explored in
Sec.~\ref{sec-lda}.

The KKR Green function calculations were done in the full-potential
mode.
\new{Only when studying the scaling of \emca\ with the SOC strength
  (Sec.~\ref{sec-lambda}), we rely on the atomic spheres approximation
  (ASA), because in that case many evaluations of \emca\ have to be
  done and the focus in that part is on trends and not so much on
  numerical values. }
The energy integrals were evaluated by contour integration on a
semicircular path within the complex energy plane, using a Gaussian
mesh of 40~points. An important convergence parameter is the maximum
angular momentum $\ell_{\text{max}}^{\text{(KKR)}}$ used for the
multipole expansion of the Green function (see Appendix \ref{sec-lmax}).  To get as accurate results as possible, we 
mostly use $\ell_{\text{max}}^{\text{(KKR)}}$=7.  However, if a
lot of calculations with different settings has to be done
(Secs.~\ref{sec-angle} and~\ref{sec-lambda}), we use
$\ell_{\text{max}}^{\text{(KKR)}}$=3, which is sufficient if the focus
is on how $E_{\text{MCA}}$ varies with the magnetization angle or with
the SOC strength and not on particular values.

The convergence of FLAPW calculations is determined by the size of the
basis.  We treated Fe $3p$, $3d$, $4s$ and Pt $5p$, $5d$, $6s$ states
as valence states and Fe $1s$, $2s$, $2p$, $3s$ and Pt $1s$, $2s$,
$2p$, $3s$, $3p$, $3d$, $4s$, $4p$, $4d$, $4f$, $5s$ states as core
states.  The expansion of the wave functions into plane waves is
controlled by the plane wave cutoff in the interstitial region. This
cutoff is specified via the product $R_{\text{MT}}K_{\text{max}}$,
where $R_{\text{MT}}$ is the smallest muffin-tin (``atomic'') sphere
radius and $K_{\text{max}}$ is the magnitude of the largest wave
vector.  We use $R_{\text{MT}}K_{\text{max}}$=8 in this study.  The
convergence of \emca\ with $R_{\text{MT}}K_{\text{max}}$ is
investigated in the Appendix \ref{sec-rk}.  The
expansion of the wave functions into atomic-like functions inside the
spheres is controlled by the angular-momentum cutoff
$\ell_{\text{max}}^{\text{(APW)}}$. We use
$\ell_{\text{max}}^{\text{(APW)}}$=10 throughout this paper.  Note
that the cutoff's $\ell_{\text{max}}^{\text{(APW)}}$ and
$\ell_{\text{max}}^{\text{(KKR)}}$ have different roles in FLAPW and
KKR-Green function methods, so their values cannot be directly
compared.

As concerns the muffin-tin radii in \wien\ calculations, the atomic
spheres are chosen so that they are smaller than the touching spheres
for the MCA energy calculations
($R_{\text{MT}}^{(\text{Fe})}$=2.2~a.u.,
$R_{\text{MT}}^{(\text{Pt})}$=2.3~a.u.,
$R_{\text{MT}}^{(\text{touch})}$=2.527~a.u.)  because in this way the
basis avoids the linearization error.  On the other hand, for
analyzing site-related magnetic moments we use touching muffin-tin
spheres because in this way we minimize the moments associated to the
interstitial region. In this way we are in a better position to
compare the \wien\ results with the \kkr\ data, where the site-related
magnetic moments are determined as moments within Voronoi polyhedra.
The stability of \emca\ with respect to $R_{\text{MT}}$'s variation is
demonstrated in the Appendix \ref{sec-rmt}.

Once the Green function components or the wave functions have been
determined, the charge density is obtained via the $\bm{k}$-space
integration over the BZ.
\new{When using the \wien\ code, the BZ integration was carried out
  using the modified tetrahedron method \cite{BJA+94}.  When using the
  \kkr\ code, the BZ integration was carried out via sampling on a
  regular $\bm{k}$-mesh , making use of the symmetry \cite{HE+02}. }
The convergence of \emca\ with respect to the the number of
$\bm{k}$-points is explored in the Appendix \ref{sec-kpts}.  Based on it, we used 800000~$\bm{k}$-points in the full
BZ for \wien\ calculations and 100000~$\bm{k}$-points in the full BZ
for \kkr\ calculations.

Considering the convergence tests as a whole, we argue that that the
numerical accuracy of our \emca\ values is about 0.1~meV for
\wien\ calculations and about 0.2~meV for \kkr\ calculations.

%%---%%---%%---%%---%%---%%---%%---%%---%%---%%---%%---%%---%%---%%

\subsection{Treatment of relativistic effects}   \label{sec-relat}

The {\sc sprkkr} code works fully relativistically, it solves a four-component
Dirac equation by default.  SOC is therefore implicitly fully included
for all states.  Nevertheless, the bare effect of the SOC can be
investigated via {\sc sprkkr} if one employes an approximate
two-component scheme \cite{EFVG96} where the SOC-related term is
identified by relying on a set of approximate radial Dirac equations.
This scheme was used recently to investigate how the MCA energy of adatoms
and monolayers on noble metals varies if SOC is selectively switched
on only at some sites \cite{SBE+14}. We employed it here for the
same purpose. 

As concerns the {\sc wien2}k code, SOC is included differently for
core and valence electrons.  The core electrons are treated fully
relativistically by solving the atomic-like Dirac equation.  For the
valence electrons the SOC is included in atomic spheres via an
approximative scheme that introduces an additional term
\begin{equation}
H_{SOC} \, = \, \xi(\bm{r}) \, \bm{L} \cdot \bm{S}
\label{Hsoc}
\end{equation}
to the spin-polarized Schr\"{o}dinger-like scalar relativistic
equation.  Technically, the 
influence of the term~(\ref{Hsoc}) is included by starting with a
scalar-relativistic FLAPW calculation without SOC. 
The eigenfunctions thus obtained are then used as a basis in which
another diagonalization is done and this time also the SOC term
Eq.~(\ref{Hsoc}) is taken into account.  This procedure is often
called second variational step \cite{MacDonald+79}. Usually this second
variational step is applied only to a subset of FLAPW eigenstates to
gain a substantial speed-up.  This subset is defined so that it
includes all scalar-relativistic eigenstates up to energy
$E_{\text{max}}$ above the Fermi level.  The $E_{\text{max}}$
parameter thus plays an analogous role as
$R_{\text{MT}}K_{\text{max}}$.  Moreover, relativistic local orbitals
($p_{1/2}$ wavefunctions) were added to the basis \cite{Kune+01}. To
achieve the highest accuracy, we set $E_{\text{max}}$ as large as
needed to include all FLAPW eigenfunctions in the second step (this
can achieved by setting $E_{\text{max}}$ of 100~Ry or higher).
More details can be found in the Appendix.  As
concerns the interstitial region, valence electrons are treated in a
non-relativistic way. In the rest of this paper
``fully relativistic calculation'' implies use of the Dirac
equation for {\sc sprkkr} and Schr\"{o}dinger equation plus separate
SOC term~(\ref{Hsoc}) in the Hamiltonian for {\sc wien}2k, unless it
is explicitly said otherwise.

The \kkr\ and \wien\ codes allow for non-relativistic and
scalar-relativistic calculations as well.  In the first case, both
valence and core electrons are treated non-relativistically.  In the
second case, the valence electrons are treated using the
scalar-relativistic approach while for the core electrons atomistic
Dirac equation is solved (this applies to both codes).

%%---%%---%%---%%---%%---%%---%%---%%---%%---%%---%%---%%---%%---%%

\subsection{Scaling of the spin-orbit coupling}   \label{sec-scale}

For a deeper understanding we want to investigate how $E_{\text{MCA}}$
depends on the SOC.  More specifically, we are interested in how
$E_{\text{MCA}}$ varies if the SOC strength is varied at the Fe and Pt
sites separately, i.e., we assume that the Hamiltonian
Eq.~(\ref{Hsoc}) can be symbolically rewritten as
\begin{equation}
H_{SOC} \, = \, \sum_{i} \lambda_{i} \, \xi_{i}(\bm{r}) \, 
    \bm{L}_{i} \cdot \bm{S}_{i}
\;\; ,
\label{eq-scale}
\end{equation}
where $\lambda_{i}$ is the scaling factor for site $i$.  Such
calculations were done via the \kkr\ code, using the approximate
scheme \cite{EFVG96} mentioned in the beginning of
Sec~\ref{sec-relat}.

%%%%%%%%%%%%%%%%%%%%%%%%%%%%%%%%%%%%%%%%%%%%%%%%%%%%

\section{Results}    \label{sec-res}

%%---%%---%%---%%---%%---%%---%%---%%---%%---%%---%%---%%---%%---%%

\subsection{Magnetic moments}  
 \label{sec-moments}

\begin{table}
\caption{ Spin magnetic moments (in $\mu_{B}$) related either to a
  FePt unit cell or just to the Fe site, for different ways of 
  including the relativistic effects.  \label{total-spin} }
\begin{ruledtabular}
\begin{tabular}{ldddd}
 & \multicolumn{2}{c}{\sc sprkkr} & \multicolumn{2}{c}{ {\sc wien2}k} \\
 & \multicolumn{1}{c}{$\mu_{\text{spin}}^{(\text{cell})}$} &
  \multicolumn{1}{c}{$\mu_{\text{spin}}^{(\text{Fe})}$} &
  \multicolumn{1}{c}{$\mu_{\text{spin}}^{(\text{cell})}$}
  & \multicolumn{1}{c}{$\mu_{\text{spin}}^{(\text{Fe})}$} \\
\hline
non relativistic     & 3.17  & 2.86  & 3.15  & 2.86  \\
scalar relativistic  & 3.21  & 2.86  & 3.21  & 2.87  \\
fully relativistic   & 3.17  & 2.83  & 3.17  & 2.84  \\
\end{tabular}
\end{ruledtabular}
\end{table}

The presence of Pt in FePt suggests that the way relativistic effects
are treated could be important.  Therefore, we calculated magnetic
moments in FePt using a non-relativistic Schr\"{o}dinger equation,
using a scalar-relativistic approach, and using a relativistic scheme.
Spin magnetic moments related either to the unit cell or only to the
Fe site are shown in Tab.~\ref{total-spin}.  We can see that
relativity has only a marginal effect on the spin magnetic moments in
FePt.  Orbital magnetic moments are more interesting in this respect
--- they would be zero in the absence of SOC.  Our results in
Table~\ref{orbital} give the orbital magnetic moment at the Fe and Pt
sites for two orientations of the magnetization.

\begin{table}
\caption{ Orbital magnetic moments (in $\mu_{B}$) related to the Fe and Pt
  atoms in FePt for magnetization either parallel to the $z$ axis
  ($\mu_{\text{orb}}^{(\bm{M} \| z)}$) or perpendicular to the
  $z$ axis ($\mu_{\text{orb}}^{(\bm{M} \| x)}$). \label{orbital} }
\begin{ruledtabular}
\begin{tabular}{cdddd}
& \multicolumn{2}{c}{Fe} & \multicolumn{2}{c}{Pt} \\ 
& \multicolumn{1}{c}{\sc sprkkr} & \multicolumn{1}{c}{{\sc wien}2k} 
& \multicolumn{1}{c}{\sc sprkkr} & \multicolumn{1}{c}{{\sc wien}2k} \\ 
\hline 
$ \mu_{\text{orb}}^{(\bm{M} \| z)}$ & 0.065 & 0.065 & 0.044 & 0.042 \\ 
$\mu_{\text{orb}}^{(\bm{M} \| x)}$  & 0.062 & 0.062 & 0.060 & 0.054 \\ 
\end{tabular}
\end{ruledtabular}
\end{table}

One can see that both codes lead to very similar values for \ms\ and
\mo.  In particular, the anisotropy of $\mu_{\text{orb}}$ at Fe and at
Pt sites is nearly the same.  Small differences between the codes in
the local magnetic moments may be due to the fact that the moments are
defined in different regions: Wigner-Seitz cells (or more precisely
Voronoi polyhedra) in {\sc sprkkr} and touching muffin-tin spheres in
\wien.  The difference would be larger if we used ``standard'' setting
of muffin-tin radii in
\wien\ ($R_{\text{MT}}^{(\text{Fe})}$=2.2~a.u.\ and
$R_{\text{MT}}^{(\text{Pt})}$=2.3~a.u.\ instead of
$R_{\text{MT}}^{(\text{Fe})}$=$R_{\text{MT}}^{(\text{Pt})}$=2.527~a.u.):
in that case, the local spin moments obtained via \wien\ would be
smaller by about 3~\% and orbital moments by about 10~\%.

%%---%%---%%---%%---%%---%%---%%---%%---%%---%%---%%---%%---%%---%%

\subsection{Magneto-crystalline anisotropy energy} 
\label{sec-MCA} 

Calculating the MCA energy by subtracting total energies for two
orientations of the magnetization as 
\begin{equation}
E_{\text{MCA}} \: \equiv \: E^{(\bm{M} \| x)} -  E^{(\bm{M} \| z)}
\label{eq-emca}
\end{equation} 
is very challenging, because the total energies and the MCA energy
differ by about eight or nine orders of magnitude. We paid a lot of
attention to the issues of convergence to get accurate numbers.  The
details can be found in the Appendix.  Here
we only mention two issues which have to be given special attention.

For full-potential {\sc sprkkr} calculations, attention has
to be paid to the multipole expansion of the Green function governed
by the cutoff $\ell_{\text{max}}^{\text{(KKR)}}$.  
\new{KKR calculations have known behavior concerning the 
  $\ell_{\text{max}}^{\text{(KKR)}}$ convergence which play role if one
  aims at high-accuracy total energy calculations
  \cite{Zel+13,AKS+14}.  Part of the problem are numerical
  difficulties connected with the evaluation of the Madelung contribution
  to the full potential for high angular momenta \cite{AWJ+11,kovzel}.
  Note that to obtain the Green function components up to
  $\ell_{\text{max}}^{\text{(KKR)}}$, one needs potential components
  up to 2$\ell_{\text{max}}^{\text{(KKR)}}$ and shape functions
  components up to 4$\ell_{\text{max}}^{\text{(KKR)}}$.  Another
  difficulty is an efficient treatment of the so-called near-field
  corrections \cite{AWJ+11,OZO+15}.  Various ways to deal with these
  issues have been suggested \cite{PAKW+11,AKS+14,OZO+15,Zel+15}.  We
  performed a test of the $\ell_{\text{max}}^{\text{(KKR)}}$
  convergence  (Appendix \ref{sec-lmax}) which
  indicate if that the $\ell_{\text{max}}^{\text{(KKR)}}$=7 cutoff is used,
  that the numerical accuracy of the MCA energy is about 0.2~meV. }

\new{For accurate MCA energy calculations using the \wien\ code, one
  has to pay special attention so that the energy parameters $E_{\ell}$
  used for calculating radial wave functions
  $u_{\ell}(r,E_{\ell})$ are determined very precisely and
  consistently.  This applies, in particular, also for the relativistic
  local orbitals.  In \wien\ this is done by searching for the
  energies where $u_{\ell}(R_{MT},E)$ changes the sign to determine
  $E_{\text{top}}$, and where it has zero slope to determine
  $E_{\text{bottom}}$. The arithmetic mean of these two energies gives
  $E_{\ell}$.  For the calculations presented here these energies had to be
  determined with an accuracy better than 0.1~mRy.  A parameter
  specific for relativistic calculations via \wien\ is
  $E_{\text{max}}$, which controls how many scalar-relativistic
  eigen-states are considered when SOC is included
  (Appendix \ref{sec-emax}).  We used $E_{\text{max}}$=100~Ry,
  meaning that all eigen-states were included. }

The MCA energy obtained by subtracting the total energies is shown in
the first line of Tab.~\ref{MCAE}.  Values obtained via {\sc sprkkr}
and \wien\ show good agreement.  Considering the convergence analysis
we performed, this allows us to state that the magnetic easy axis of
FePt is out-of-plane and the MCA energy is 3.0~meV within the LDA
framework (for the VWN exchange-correlation functional).

\begin{table}
\caption {MCA energy $E_{\text{max}}$ of FePt (in meV) calculated by
  two   approaches.  \label{MCAE} }
\begin{ruledtabular}
\begin{tabular}{ldd}
 & \multicolumn{1}{c}{\sc sprkkr} & \multicolumn{1}{c}{{\sc wien}2k}  \\
 \hline 
subtracting total energies  &  3.04  &   2.99  \\ 
magnetic force theorem      &  3.12  &   2.85  \\ 
\end{tabular}
\end{ruledtabular}
\end {table}

Obtaining the MCA energy by subtracting the total energies is
computationally very costly.  The need for self-consistent
calculations for two magnetization directions can be avoided if one
relies on the magnetic force theorem. In this approach the MCA energy
is calculated using a frozen spin-dependent potential
\cite{Mackinttosh+80,WWW+96}. The MCA energy is then obtained either
by subtracting the band-energies or by evaluating the torque at
magnetization tilt angle of 45$^{\circ}$ \cite{Wang+96,SBE+14}.  As
the magnetic force theorem is frequently employed, we applied it here
as well.  The results are shown in the second line of Tab.~\ref{MCAE}.
We can see that the magnetic force theorem yields very similar values
as if total energies are subtracted.

%%---%%---%%---%%---%%---%%---%%---%%---%%---%%---%%---%%---%%---%%

\subsubsection*{Relation between \emca\ and anisotropy of \mo}

For the sake of completeness we checked also the Bruno formula
\cite{Bruno+89}, which links the MCA energy to the anisotropy of
orbital magnetic moment.  The Bruno formula \cite{Bruno+89} (as well
as the slightly more sophisticated van der Laan formula \cite{vdL+98})
can be derived from second order perturbation theory if some
additional assumptions are made.  It is often employed in the context
of x-ray magnetic circular dichroism experiments that give access to
the anisotropy of orbital magnetic moment via the so-called sum rules.

\new{Even though the formula was originally derived for systems with
  only one atomic type, the relation between the MCA energy and the
  anisotropy of orbital magnetic moments has been frequently applied
  also for multicomponent systems
  \cite{RKF+01,GC+12,AMV+12,MOK+13,KMS+13,USI+15}.  In such a case an
  estimate of \emca\ can be made by evaluating (cf.~Ravindran
  \ea\ \cite{RKF+01} and Andersson \ea\ \cite{ASE+07})
\begin{equation}
E_{\text{MCA}} \, = \, \sum_{i} \frac{\xi_{i}}{4} \, 
\left(\mu_{\text{orb}}^{(i, \bm{M} \| z)} - \mu_{\text{orb}}^{(i,
  \bm{M} \| x)} \right) 
\, ,
\label{bruno}
\end{equation}
where $i$ labels the constituting atoms.  This equation is valid only if
off-site spin-flip terms are neglected \cite{RKF+01,ASE+07,KS+12}. }

We evaluated Eq.~(\ref{bruno}) using SOC parameters
$\xi^{(\text{Fe})}$=65~meV and $\xi^{(\text{Pt})}$=712~meV, as
obtained from ab-initio calculations for FePt relying on the method
described by Davenport \ea\ \cite{DWW+88}.  We obtained
$E_{\text{MCA}}=-2.62$~meV using \kkr\ results and
$E_{\text{MCA}}=-2.09$~meV using \wien\ results.  The sign of
\emca\ evaluated from Eq.~(\ref{bruno}) is wrong, indicating that
this formula does not provide a suitable framework for studying the
MCA of FePt.  Technically, the reversal of the sign of \emca\ obtained
via Eq.~(\ref{bruno}) is due to \mo\ at Pt (see Tab.~\ref{orbital}):
we have \mm{\mu_{\text{orb}}^{(\bm{M} \| z)} >
  \mu_{\text{orb}}^{(\bm{M} \| x)}}\ at the Fe site and
\mm{\mu_{\text{orb}}^{(\bm{M} \| x)} > \mu_{\text{orb}}^{(\bm{M} \|
    z)}} at the Pt site.  As $\xi^{(\text{Pt})}$ is much larger than
$\xi^{(\text{Fe})}$, the Pt-related term dominates in
Eq.~(\ref{bruno}).

The failure of the Bruno formula~(\ref{bruno}) does not automatically
imply that \sopt\ cannot be used for describing the MCA of FePt.
Namely, it is likely that additional assumptions employed in the
derivation of Eq.~(\ref{bruno}) are not fulfilled; 
\new{in particular, for Pt atoms, the exchange splitting and SOC will
  be of the same order of magnitude. }
Two more indicative tests whether \sopt\ \new{itself } provides a
good framework for understanding the MCA of FePt are presented below.

%%---%%---%%---%%---%%---%%---%%---%%---%%---%%---%%---%%---%%---%%

\subsection{Dependence of the total energy on the orientation of the
  magnetization axis}

\label{sec-angle}

Accurate calculations can provide information on the full 
\new{form  of the } 
dependence of the total energy on the angle $\theta$ between the
magnetization direction and the $z$ axis.  For tetragonal systems the
first two terms in the directional cosines expansion of the total
energy are
\begin{equation}
\label{angle}
E(\theta) \, - \, E_{0} \: = \: 
K_{1} \, \sin^{2}\theta \, + \, K_{2} \, \sin^{4}\theta 
\; .
\end{equation}
Here we omit the azimuthal dependence, keeping $\phi$=0$^{\circ}$.  If the
influence of SOC is included via the explicit term Eq.~(\ref{Hsoc}),
then application of second order perturbation theory leads to a
\new{simple } 
dependence of the total energy on the angle $\theta$ as
\[
E(\theta) \, - \, E_{0} \: = \: 
K_{1} \, \sin^{2}\theta 
\; ,
\]
meaning that only the first term survives in Eq.~(\ref{angle})
\cite{Bruno+89,ABS+06}. Inspecting the full $E(\theta)$ dependence as
obtained via fully-relativistic ab-initio calculations thus provides
the possibility to estimate to what degree a treatment of MCA based on
second order perturbation theory is adequate: large $K_{2}$
coefficient implies large deviations from second order perturbation
theory.

\begin{figure}
\includegraphics[viewport=0.5cm 0.5cm 9.1cm 6.8cm]{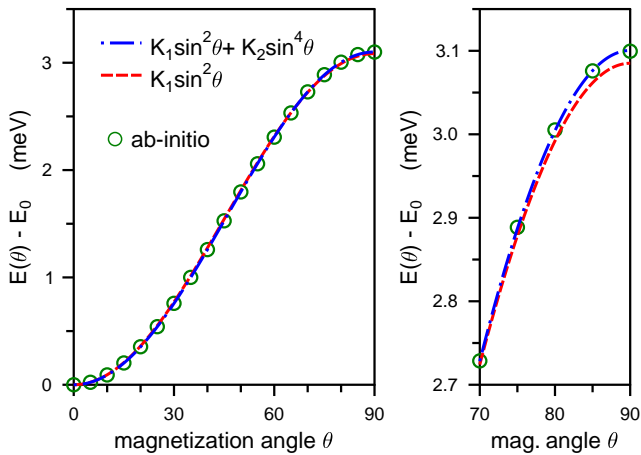}
\caption{Dependence of the total energy on the magnetization angle
  $\theta$ (circles) and its fit either as
  \mm{K_{1}\sin^{2}\theta}\ (dashed line) or as
  \mm{K_{1}\sin^{2}\theta+K_{2}\sin^{4}\theta} (dash-dotted line).  An
  overall view is in the left panel, a detailed view on the region
  close to $\theta$=90$^{\circ}$ is in the right panel.
\label{etheta} }
\end{figure}

We performed a series of calculations for different magnetization tilt
angle~$\theta$, using the \kkr\ code.  The MCA energy was evaluated as
a difference of total energies.  The results are shown via circles in
Fig.~\ref{etheta}.  Because we wanted to have a fine $\theta$-mesh, we
had to perform a lot of calculations; therefore, we used
$\ell_{\text{max}}^{\text{(KKR)}}$=3 in this section.  The numerical
value for $\theta$=90$^{\circ}$ thus differs a bit from
Tab.~\ref{MCAE}, where the \new{$\ell_{\text{max}}^{\text{(KKR)}}$=7}
cutoff was used.

The ab-initio data were fitted via Eq.~(\ref{angle}).  If only the
\mm{K_{1}\sin^{2}\theta}\ term is employed (taking $K_{2}$=0), we
obtain $K_{1}$=3.085~meV.  If both terms in Eq.~(\ref{angle}) are
employed, we obtain $K_{1}$=3.008~meV and $K_{2}$=0.092~meV.  Even
though both fits look nearly the same in the overall view, a detailed
analysis shows that the fit with both terms is significantly better
(cf.~the right panel in Fig.~\ref{etheta}).  Using even higher order
terms in the fit did not lead to a significant improvement.  

\new{To summarize, our calculations show that the dependence of the
  total energy on the magnetization angle is fully described by
  Eq.~(\ref{angle}). The ratio of the coefficients $K_{2}/K_{1}$ is
  0.03, thus we deduce that the MCA of FePt is dominated by the 
  \sopt\ but there is also a small but identifiable
  contribution which cannot be described by it. }

%%---%%---%%---%%---%%---%%---%%---%%---%%---%%---%%---%%---%%---%%

\subsection{Dependence of the MCA energy on spin orbit
  coupling}

\label{sec-lambda}

\begin{figure}
\includegraphics[viewport=0.5cm 0.5cm 7.0cm 4.9cm]{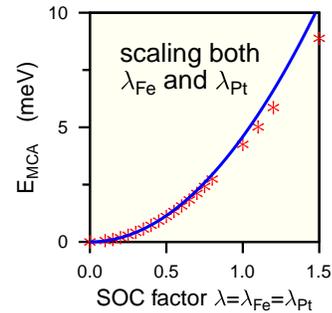}
\caption{Dependence of $E_{\text{MCA}}$ on the SOC scaling factor
  $\lambda$.  The markers denote calculated values of
  $E_{\text{MCA}}$, the line represents a fit to these data within the
  $\lambda \in [0;0.4]$ interval.
  \label{fig-scal-both} }
\end{figure}

If the \mca\ is described within \sopt, it scales with the square of
the SOC-scaling parameter $\lambda$, $E_{\text{MCA}}\sim\lambda^{2}$
\cite{Bruno+89,vdL+98,ABS+06}. Inspecting the
$E_{\text{MCA}}(\lambda)$ dependence thus provides another criterion
to what degree \sopt\ is sufficient to describe \mca\ of FePt.
\new{To get type-specific information, one should scale \lfe\ and
  \lpt\ separately.  In that case, however, the scaling of \emca\ with
  SOC takes a somewhat more complicated form \cite{ASE+07}
\begin{equation}
E_{\text{MCA}}(\lambda_{\text{Fe}},\lambda_{\text{Pt}}) 
 \; = \;  A \, \lambda_{\text{Fe}}^{2} 
 \: + \:  B  \, \lambda_{\text{Fe}} \, \lambda_{\text{Pt}}
 \: + \:  C \, \lambda_{\text{Pt}}^{2}
\; .
\label{eq-double}
\end{equation}
The scaling of \emca\ with SOC will thus retain a quadratic
form only if the scaling is uniform (\lfe=\lpt) or if SOC for one of
the atomic types is zero (recovering thus the case of a
single-component system \cite{Bruno+89,vdL+98,ABS+06}). }

We start by calculating $E_{\text{MCA}}$ for a uniform SOC scaling,
i.e., \lfe=\lpt.  We vary $\lambda$ from 0 to 1.5 to cover the
non-relativistic as well as the relativistic regime: if $\lambda$ is
zero, there is no spin orbit coupling, if $\lambda$ is 1, we recover
the standard relativistic case.  The calculations were done with the
\kkr\ code, employing the scheme described in Sec.~\ref{sec-scale} and
evaluating \emca\ by subtracting total energies.  To reduce the
computer requirements, we performed all the calculations in
this section with $\ell_{\text{max}}^{\text{(KKR)}}$=3 in the ASA
mode; this enables us to use a fine $\lambda$ mesh so that the curve
fitting is reliable.  The results are shown by points in
Fig.~\ref{fig-scal-both}.  Employment of the ASA obviously leads to
less acurate results than for full-potential calculations:
\emca\ obtained within the ASA is by about 1~eV larger than
\emca\ obtained for full potential.  However, this does not affect our
conclusions concerning the scaling of \emca\ with strength of the SOC.

To verify the predictions of the perturbation theory, we 
fit calculated $E_{\text{MCA}}(\lambda)$ with the quadratic function, 
\begin{equation}
E_{\text{MCA}}(\lambda) \; = \; a \, \lambda^{2}
\; .
\label{eq-fit}
\end{equation}
Perturbation theory should work well for small values of $\lambda$
while it can be less appropriate for large values of $\lambda$.  So
the fit to the function (\ref{eq-fit}) is performed in such a way that
the $a$ coefficient is sought only for $\lambda$ in the range between
zero and 0.4 (the upper value was arbitrarily chosen just for
convenience).  One can see from Fig.~\ref{fig-scal-both} that while
the fit describes the ab-initio data very well within the \mm{\lambda
  \in [0;0.4]}\ range, there are small but clear deviations for larger
$\lambda$.  This suggests that while \sopt\ accounts for the dominant
mechanism of \mca\ of FePt, some effects beyond it are also 
\new{present. }

\begin{figure}
\includegraphics[viewport=0.5cm 0.5cm 9.0cm 4.9cm]{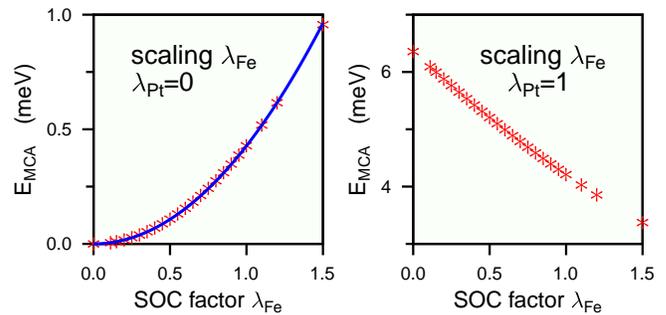}
\caption{Dependence of $E_{\text{MCA}}$ on the SOC scaling factor at
  the Fe sites \lfe.  The markers denote calculated values of
  $E_{\text{MCA}}$, the line in the left panel represents a fit to
  these data within the $\lambda_{\text{Fe}} \in [0;0.4]$ interval.
\label{fig-scal-fe} }
\end{figure}

\begin{figure}
\includegraphics[viewport=0.5cm 0.5cm 9.0cm 4.9cm]{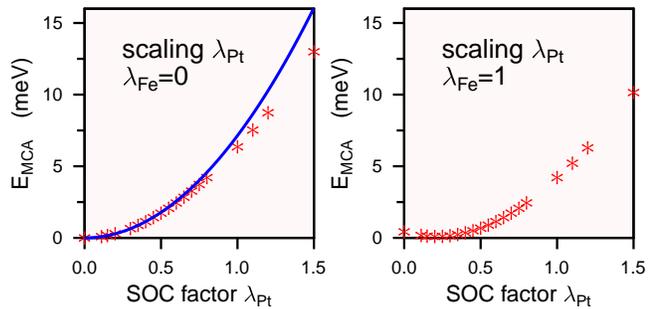}
\caption{Dependence of $E_{\text{MCA}}$ on the SOC scaling factor at
  the Pt sites \lpt.  The markers denote calculated values of
  $E_{\text{MCA}}$, the line in the left panel represents a fit to
  these data within the $\lambda_{\text{Pt}} \in [0;0.4]$ interval.
\label{fig-scal-pt}}
\end{figure}

To learn more \new{about atom-specific contributions to MCA, } let us
scale the SOC at the Fe and Pt sites separately.  When varying
\lfe\ or \lpt\ we further distinguish two cases --- either the SOC at
the remaining species is totally suppressed ($\lambda$=0) or it is
kept at its ``normal'' value ($\lambda$=1). Results for scaling SOC at
the Fe sites are shown in Fig.~\ref{fig-scal-fe}, results for scaling
SOC at the Pt sites are shown in Fig.~\ref{fig-scal-pt}.  Fits to the
quadratic dependence of \emca\ on \lfe\ or on \lpt\ were done only in
case that SOC at the other site is suppressed. 
\new{Namely, if $\lambda$ at the other atomic type is non-zero, the
  functional dependence is more complicated --- see
  Eq.~(\ref{eq-double}) --- and fitting $E_{\text{MCA}}(\lambda)$ with
  the simple Eq.~(\ref{eq-fit}) would not make sense.}
Similarly as in the case of the uniform scaling, the fits were
attempted for $\lambda$ in the [0;0.4] interval.

Concerning the case when SOC is varied at the Fe sites, one can see
that if \lpt=0, the dependence of $E_{\text{MCA}}$ on \lfe\ is
perfectly accounted for by \sopt: the quadratic fit describes the
$E_{\text{MCA}}(\lambda_{\text{Fe}})$ dependence very well also
outside the [0;0.4] interval in which the $a$ coefficient was sought
(left graph in Fig.~\ref{fig-scal-fe}).  This suggests that it must be
the strong SOC at Pt sites which makes the $E_{\text{MCA}}(\lambda)$
curve in Fig.~\ref{fig-scal-both} to deviate from a perfect parabola.
Indeed, if SOC at Pt sites is switched on (right graph in
Fig.~\ref{fig-scal-fe}),
\new{the $E_{\text{MCA}}(\lambda)$ functional dependence changes
completely. }

Let us turn now to the case of varying \lpt.  If there is no SOC at
the Fe sites, the $E_{\text{MCA}}(\lambda_{\text{Pt}})$ dependence is
described by the fitted parabola only for low values of \lpt\ (left
graph in Fig.~\ref{fig-scal-pt}).  If \lpt\ increases beyond the
fitting interval of [0;0.4], deviations of ab-initio data points from
the fit by Eq.~(\ref{eq-fit}) are similar as for uniform SOC fit
presented in Fig.~\ref{fig-scal-both}. So it follows from our analysis
that the effect of SOC at the Fe sites can be accounted for by
\sopt\ while the effect of SOC at the Pt sites goes beyond it.

%%---%%---%%---%%---%%---%%---%%---%%---%%---%%---%%---%%---%%---%%

\subsection{Dependence of the MCA energy on the LDA 
  \xcf}

\label{sec-lda}

Usually the calculated properties of solids do not crucially depend on
which form of the LDA \xcf\ is used.  However, as the MCA energy is a
very sensitive quantity, it is useful to investigate how the
\emca\ varies if different LDA exchange-correlation functionals are
used.  Apart from the VWN \xcf\ used throughout this work we include
in the comparison the Perdew and Wang \xcf\ \cite{Perdew+92} (the
default for \wien) and functionals suggested by von Barth and Hedin
\cite{Barth+72} and by Moruzzi, Janak and Williams \cite{MJW+78}.

We evaluated \emca\ by subtracting total energies for this
test.  The results are summarized in Tab.~\ref{XC-func}.  One can see
that different LDA functionals lead to MCA energies that differ from
each other by 0.1--0.2~meV.

\begin{table}
\caption{The MCA energy of FePt (in meV) calculated by subtracting
  total energies for different exchange and correlation
  functionals.  \label{XC-func} }
\begin{ruledtabular}
\begin{tabular}{ldd}
 & \multicolumn{1}{c}{\sc sprkkr} & \multicolumn{1}{c}{\sc wien}2k \\
 \hline 
Vosko and Wilk and Nusair \cite{VWN+80}  &  3.04 &  2.99  \\ 
Perdew and Wang \cite{Perdew+92}              &
\multicolumn{1}{c}{---}   
                                                 &  3.02   \\ 
von Barth and Hedin \cite{Barth+72}           &  3.29 &  3.18  \\ 
Moruzzi, Janak and Williams \cite{MJW+78}     &  2.97  &   
                                       \multicolumn{1}{c}{---}     \\ 

\end{tabular}
\end{ruledtabular}
\end {table}

%%---%%---%%---%%---%%---%%---%%---%%---%%---%%---%%---%%---%%---%%

\subsection{Relativistic effects in the density of states}
\label{sec-dos}

\begin{figure}
\centering
  \begin{tabular}{@{}cc@{}}
\includegraphics [width=0.5\linewidth]{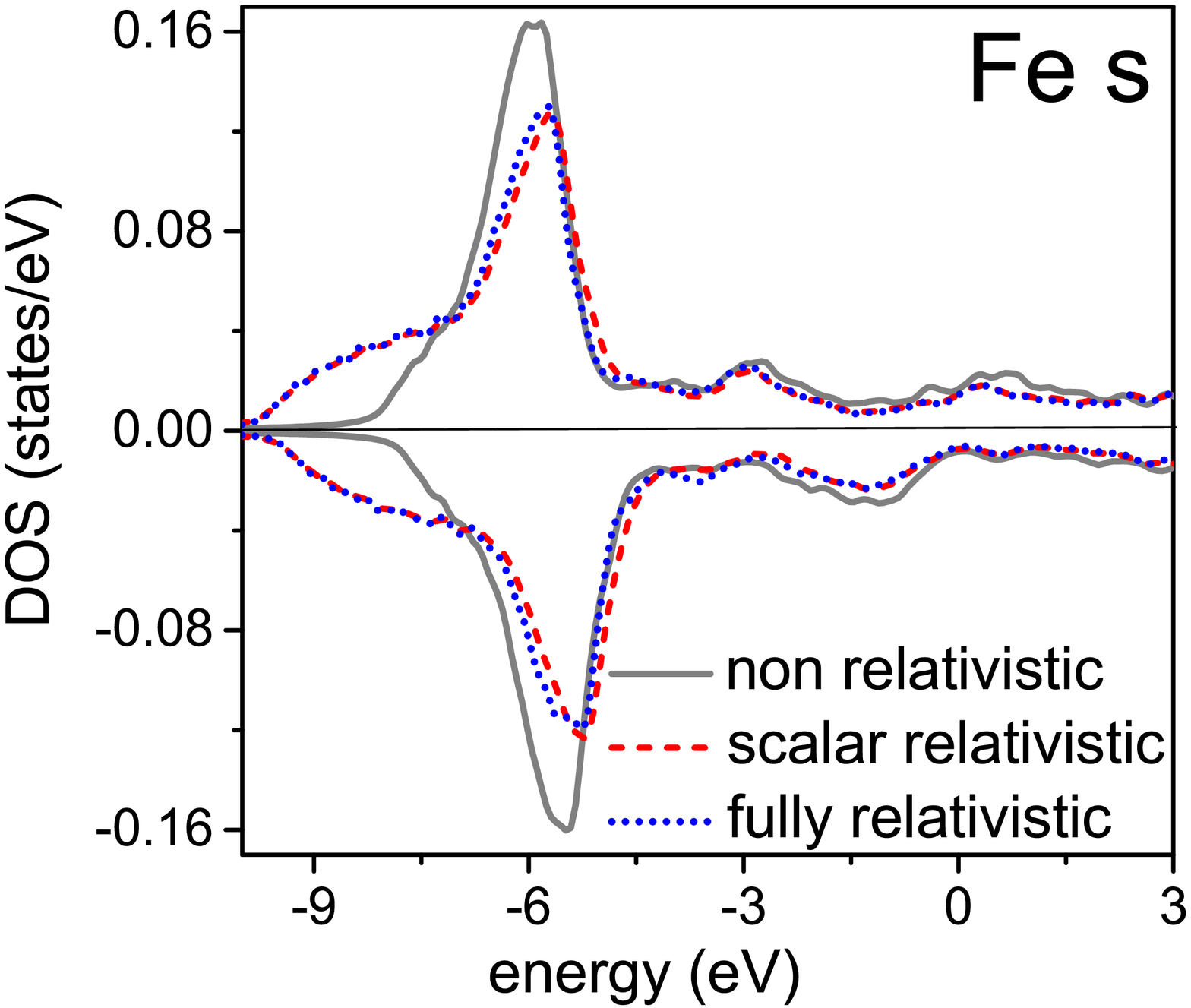} &
\includegraphics [width=0.5\linewidth]{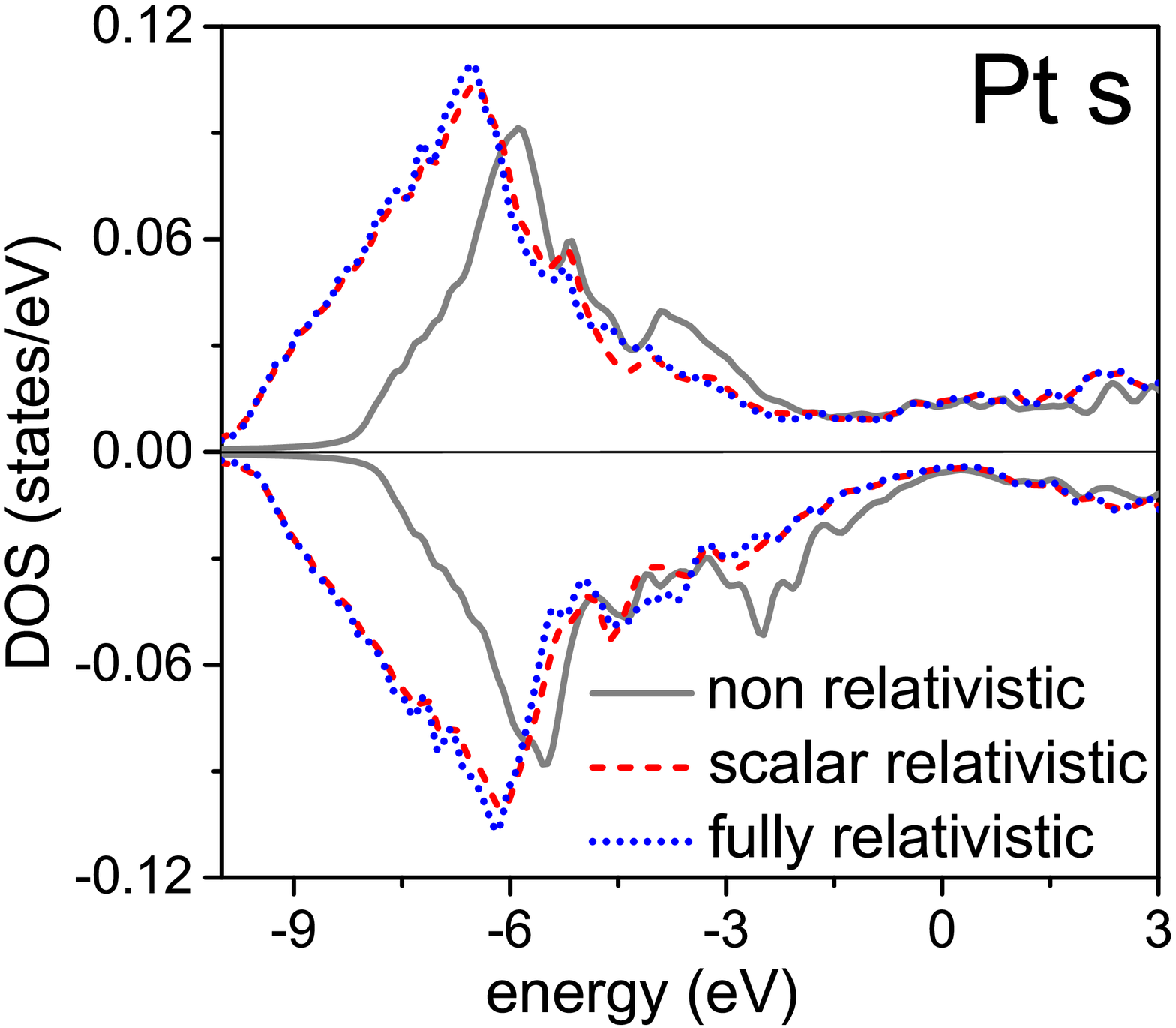}\\
\includegraphics [width=0.5\linewidth]{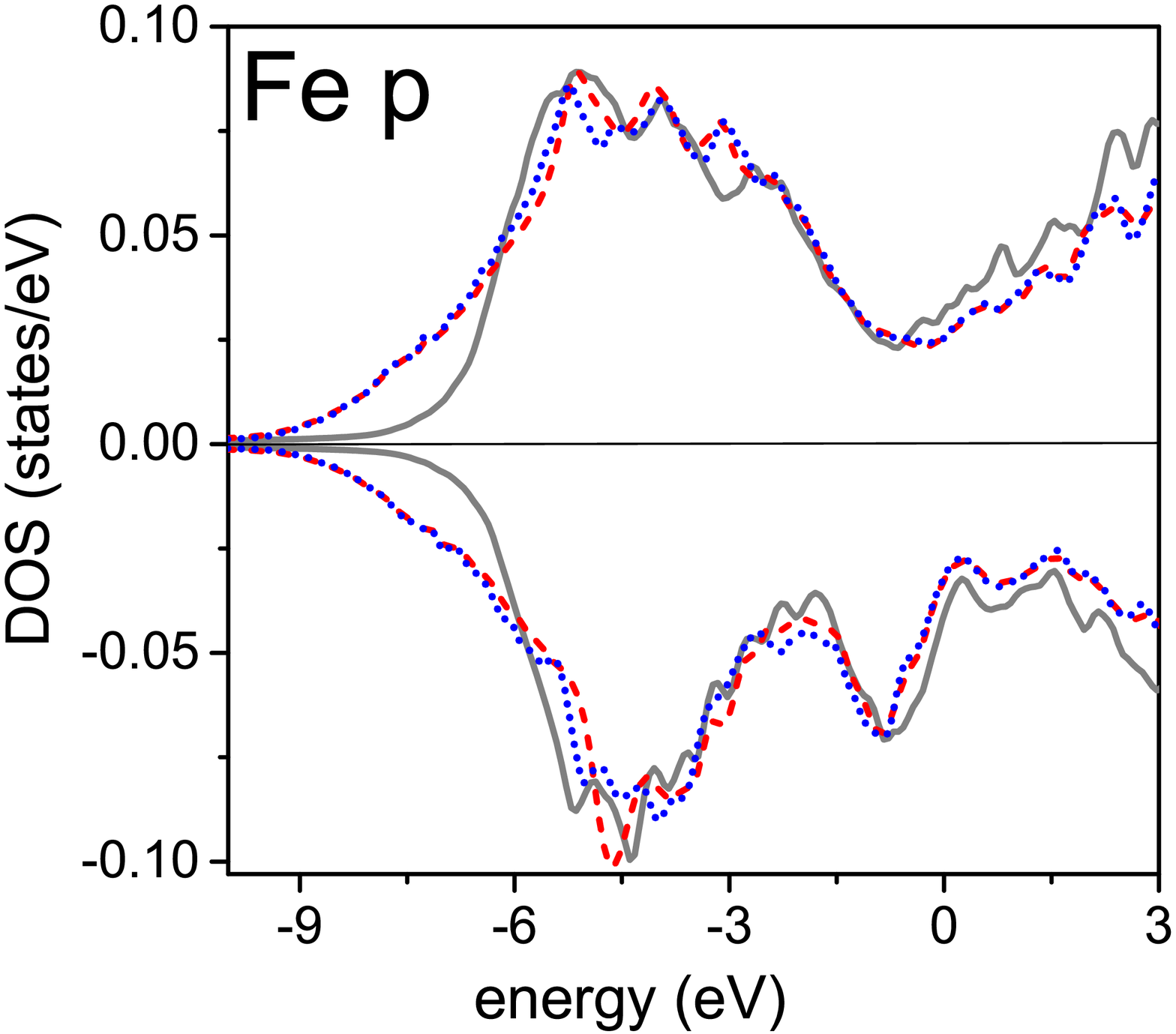}&
\includegraphics [width=0.5\linewidth]{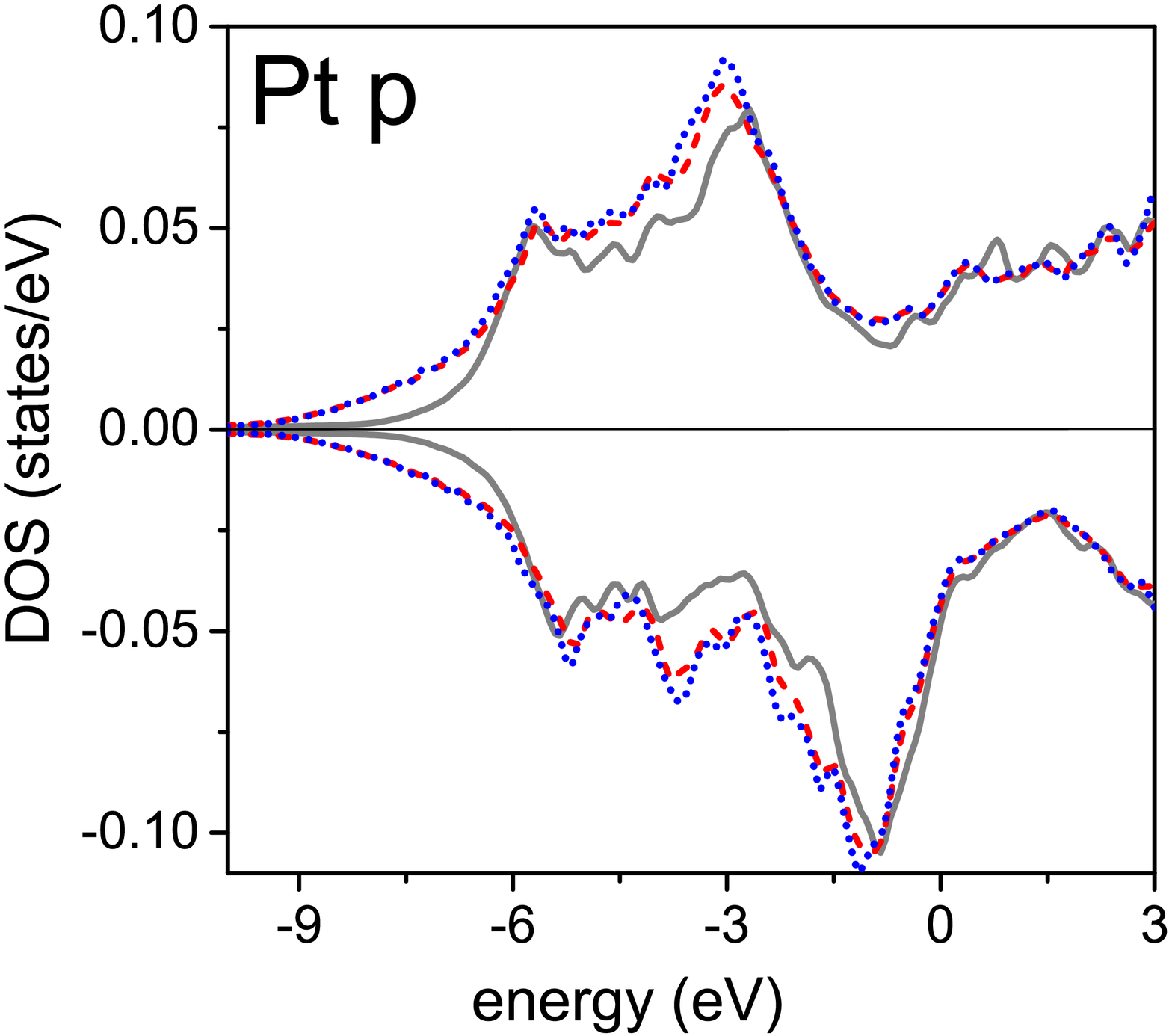}\\
\includegraphics [width=0.5\linewidth]{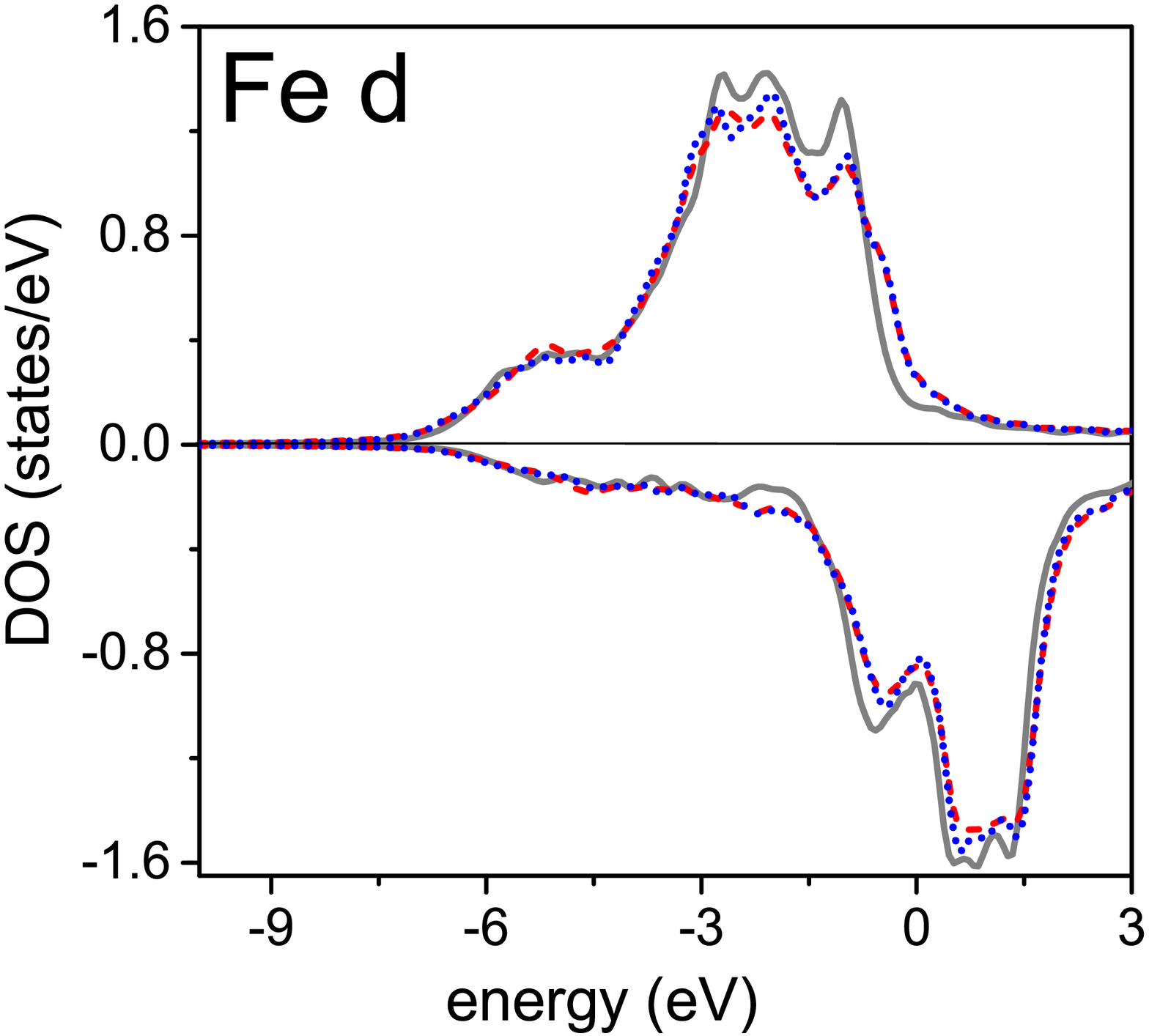}&
\includegraphics [width=0.5\linewidth]{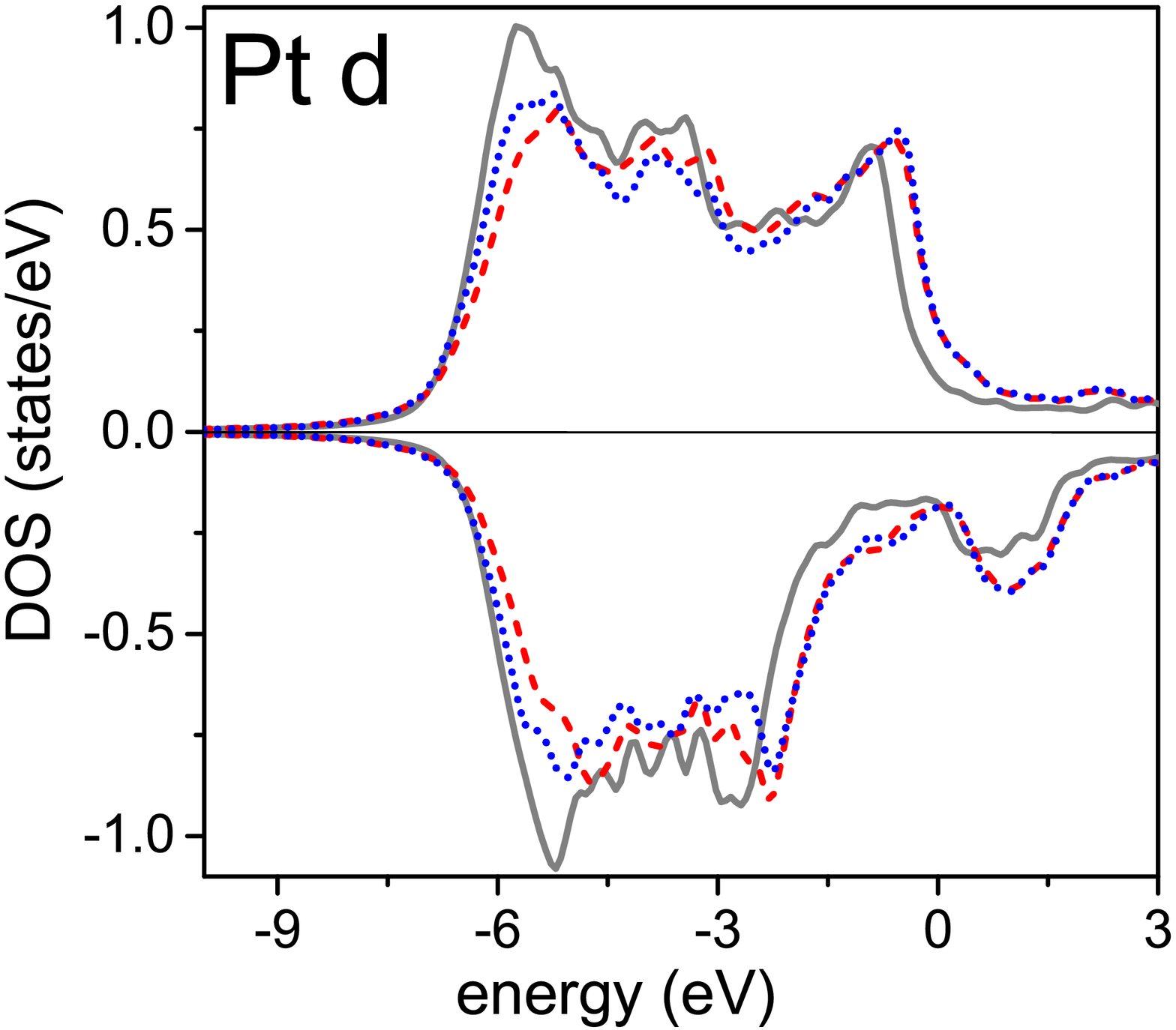}\\
\end{tabular}
\caption{Partial spin-resolved angular-momentum-projected density of states for Fe 
and Pt sites calculated within a non-relativistic, a scalar-relativistic 
and a fully-relativistic framework. \label{DOS}}
\end{figure}

Fig.~\ref{DOS} depicts the influence of relativity on the density of
states (DOS) resolved in angular momentum components respective to Fe
and Pt sites.  The data presented here were obtained using the
\kkr\ code; data obtained using the \wien\ code look practically the
same.

Generally, there is a significant change in the DOS when going from
non-relativistic to scalar-relativistic case and only a minor change
when going from scalar-relativistic to the fully relativistic case.
The largest difference between non-relativistic and relativistic case
is for the $s$~states.  This may be due to the fact that $s$~electrons
have a large probability density near the nucleus where relativistic
effects (mass-velocity and Darwin term) are stronger than at larger
distances.  Largest difference between scalar relativistic and fully
relativistic calculations are for the Pt~$d$~states, where also the
SOC is expected to be stronger than for the other cases.

For Pt $s$ and $d$ states one can make an interesting comparison with
atomic results for Au \cite{PD+79} which are often quoted when
relativistic effects in solids are discussed.  It follows from
Fig.~\ref{DOS} that relativistic effects shift valence Pt $6s$~states
to lower energies due to the orthogonality constrains to the more
localized $1s$ state and Pt $5d$~states to higher energies due to a
better screening of the nucleus by innermost electrons. The same
happens for 6$s$ and 5$d$ atomic states of Au, respectively. So we can
infer that the mechanism through which relativity affects Pt states is
essentially atomic-like and common to all 5$d$ noble metals.

%%%%%%%%%%%%%%%%%%%%%%%%%%%%%%%%%%%%%%%%%%%%%%%%%%%%

\section{Discussion} 

\new{Our aim was to get realiable quantitative information on the MCA
  of FePt, which we take as an archetypal layered system of magnetic
  and non-magnetic transition metals.  We employed two quite different
  computational procedures. }  
Both of them yield similar values for the MCA energy.
\new{Numerical stability of results is well documented by convergence
  tests presented in the Appendix. }
Therefore the results can be trusted to represent the true LDA value
of the MCA energy.  Our data can be used as a benchmark for LDA
calculations.

\new{Relativistic effects are implemented in the \wien\ code in an
  approximative way, accounting for the SOC by a separate term (see
  Eq.~(\ref{Hsoc})) which is added to the scalar-relativistic
  Hamiltonian.  Most codes rely on this approach when they deal with
  SOC.  The \kkr\ scheme, on the other hand, solves the Dirac equation
  so it does not use approximations when dealing with relativistic
  effects. }
Good agreement between MCA energies obtained via the \wien\ code
and via the \kkr\ code shows that dealing with relativity by invoking
the separate term Eq.~(\ref{Hsoc}) is justified in our case.  
\new{As we are studying FePt, i.e., a compound containing an element
  with a strong SOC, it is likely that the approximative scheme
  associated with Eq.~(\ref{Hsoc}) is sufficiently accurate for most
  common situations and/or systems. }
One should only make sure that a sufficiently large basis for the
second variation step is taken (see Appendix).

\new{We calculated \emca\ both via subtracting total energies and via
  the magnetic force theorem.  Using the magnetic force theorem is
  technically much more convenient than subtracting total energies.
  Knowing limits of its reliability it thus vital.  For pure Fe
  monolayers the magnetic force theorem was shown to be valid to a
  high accuracy \cite{NSF+09,LSB+13}.  However, there are indications
  that this may no longer be true for systems with normally
  non-magnetic atoms with large induced moments and strong SOC
  \cite{BH+09,BLD+10}.  For such atoms one would expect rather large
  changes of the spin-polarized electron density upon rotation of the
  magnetization.  This applies also for the Pt atoms in FePt.  Our
  results indicate, nevertheless, that the magnetic force theorem
  yields quite accurate values for \emca\ for FePt (Tab.~\ref{MCAE}).
  One can conjecture that this would be the case for similar layered
  systems as well. }

When comparing our \emca\ with experiment (1.3--1.4~meV)
\cite{Ivanov+73}, it is evident that the LDA result does not 
quite agree with it. Clearly one has to go beyond LDA 
\new{for a quantitative description of }
MCA of FePt.  It does not matter in this respect which specific form
of the LDA functional is used.  Nevertheless, as different LDA
functionals lead to similar but still visibly different values of
\emca\ (cf.\ Tab.~\ref{XC-func}), each calculation of the MCA energy
should be always accompanied by information which parametrization of
the LDA functional was employed.

Employing the generalized gradient approximation (GGA) does not lead
to substantial improvement with respect to the LDA.  We obtained
$E_{\text{MCA}}=2.73$~meV for the frequently used PBE-GGA form
\cite{Perdew+96} (using the \wien\ code and evaluating the MCA energy
as a difference of total energies).  It is worth to note in this
respect that Shick and Mryasov were able to obtain the MCA energy of
FePt as 1.3~meV by using the LDA+$U$ approach and searching for
suitable site-related values of the $U$ parameter \cite{Shick+03}.
Interestingly, if many-body effects are described via the orbital
polarization term of Brooks \cite{Brooks+85}, calculated
$E_{\text{MCA}}$ is not significantly improved in comparison with the
LDA \cite{Shick+03,RKF+01,Solovyev+95,Oppeneer+98} --- despite the
fact that this approach proved to be useful when calculating orbital
magnetic moments of transition metals \cite{Eriksson+90,Hjortstam+96}.

The Bruno formula, derived  originally for single-component systems
only, has recently been  employed also for systems
where there is more than one magnetic element
\cite{RKF+01,GC+12,AMV+12,KMS+13}. In our case the Bruno formula
suggests a wrong magnetic easy axis, hence it not a suitable tool for
understanding the MCA of FePt.  Similar observations were made
earlier for other compounds containing 3$d$ and 5$d$ elements
\cite{ASE+07,SMME+08,KS+12}, so we suggest that intuition based on
analysis of orbital moments should not be used for these systems
\new{--- despite its appeal and success in monoelemental systems. }

Concerning a more detailed view on the mechanism of MCA, we found that
even though MCA of FePt is dominated by a \sopt\ mechanism (as found
earlier by Kosugi \ea\ \cite{KMI+14} by analyzing the dependence of
\emca\ of FePt on $c$/$a$), effects beyond it are clearly present as well.
These effects could be identified (i) by analyzing the full angular
dependence of the total energy and (ii) by inspecting how the MCA
energy depends on the SOC strength.  Separate scaling of SOC at Fe and
Pt sites allows us to deduce that the deviations from a pure
\sopt\ mechanism have their origin at the Pt sites.  
\new{One possible mechanism that is beyond the standard \sopt\ is reoccupation
  of states close to the Fermi level \cite{Daalderop+91,SMP+16}. }

\new{Another implication comming from our analysis of the full
  angular dependence of the total energy is that one can indeed use
  the torque implemenetation of the magnetic force theorem: replacing
  the difference of energies \mm{E(90^{\circ})-E(0^{\circ})}\ by the
  torque at $45^{\circ}$ can be done only if Eq.~(\ref{angle}) is
  valid \cite{Wang+96,SBE+14}.  It follows from the results shown in
  Fig.~\ref{etheta} that this indeed is the case. }

%%%%%%%%%%%%%%%%%%%%%%%%%%%%%%%%%%%%%%%%%%%%%%%%%%%%

\section{Conclusions} 

If electronic structure calculations performed by means of FLAPW and
KKR methods are properly converged, they yield the same results even
for such sensitive quantities as the \mca\ energy. The proper
LDA value of the MCA energy for FePt (\new{3.0~meV} for the VWN \xcf) is
significantly larger than in experiment (1.3~meV), meaning that the
MCA of FePt can be described properly only if many-body effects beyond
the LDA are included.  As our value of \emca\ was obtained by two
different methods and the convergence of both of them was carefully
checked, it can be used as a benchmark in future calculations.

It is not really important whether relativistic effects for FePt are
accounted for by solving the full Dirac equation or whether the
spin-orbit coupling is treated as a correction to the
scalar-relativistic Hamiltonian.  The main mechanism of MCA in FePt
can be described within the framework of \sopt\ but a significant
contribution not accountable for by the \sopt\ is present as well.

%%%%%%%%%%%%%%%%%%%%%%%%%%%%%%%%%%

\begin{acknowledgments}
We would like to acknowledge CENTEM project (CZ.1.05/2.1.00/03.0088), CENTEM PLUS (LO1402) and COST CZ LD15147 of The Ministry of 
Education, Youth and Sports (Czech Republic). Computational time has been provided with
the MetaCentrum (LM205) and CERIT-SC (CZ.1.05/3.2.00/08.0144)
infrastructures.  In addition we would like to thank for travel
support from EU-COST action MP1306 (EUSpec). 
\end{acknowledgments}

%%%%%%%%%%%%%%%%%%%%%%%%%%%%555555%%%%
%%%%%%%%%%%%%%%%%%%%%%%%%%%%%%%
\appendix*
\label{appendix}

\section{Convergence tests}
The total energies and the MCA energy can differ by about eight or
nine orders of magnitude.  Very well converged calculations are thus
required for precise values of the MCA energy.  We checked the
influence of different technical parameters on the MCA energy if the
{\sc wien}2k and {\sc sprkkr} codes are used. Some results which may
be interesting for those practicing such calculations are presented in
this appendix.

The \emca\ values presented in this appendix sometimes
differ from the values presented in the Results section of the main
paper.  This is because in order to save computer resources, when
studying the dependence of \emca\ on a particular convergence
parameter, the other parameters were sometimes set to lower values
than what would lead to the most accurate results.  These
circumstances do not influence the outcome of the convergence tests.

Unless explicitly stated otherwise, the setting of technical
parameters in this appendix is the following
(cf.~Sec.~\ref{sec-dets}): $\ell_{\text{max}}^{\text{(KKR)}}$=3 (for
\kkr), $R_{\text{MT}}^{(\text{Fe})}$=2.2~a.u.,
$R_{\text{MT}}^{(\text{Pt})}$=2.3~a.u.,
$R_{\text{MT}}K_{\text{max}}$=8,
$\ell_{\text{max}}^{\text{(APW)}}$=10, $E_{\text{max}}$=100~Ry (for
\wien).  Reciprocal space integrals were evaluated using a mesh of
100000~$\bm{k}$-points in the full BZ (both codes). 

Based on the convergence tests presetend here, we argue that that the
numerical accuracy of \emca\ values presented in the main paper is
about 0.1~meV for \wien\ calculations and about 0.2~meV for
\kkr\ calculations.

%%---%%---%%---%%---%%---%%---%%---%%---%%---%%---%%---%%---%%---%%

\subsection{Convergence of \kkr\ calculations with
  $\ell_{\text{max}}^{\text{(KKR)}}$ } 

\label{sec-lmax}

KKR calculations of total energies are quite sensitive to the
$\ell_{\text{max}}^{\text{(KKR)}}$ cutoff. Therefore, we explore the
dependence of our results on this parameter.  The results are shown in
Tab.~\ref{lmax}.  It follows from the table that cutting the angular
momentum expansion at $\ell_{\text{max}}^{\text{(KKR)}}$=3 (as it is
commonly done for transition metals) yields qualitatively correct
value for the MCA energy.

One can see from Tab.~\ref{lmax} that even for
  $\ell_{\text{max}}^{\text{(KKR)}}=7$, a full convergence still has
  not been reached.  However, increasing
  $\ell_{\text{max}}^{\text{(KKR)}}$ further would be computationally
  very demanding and, moreover, the issue of
  $\ell_{\text{max}}^{\text{(KKR)}}$ convergence would get
  intertwinned with numerical problems in evaluating the Madelung
  potential and near-field corrections, so the real benefit of it
  would be dubious.  We conclude that this limits the numerical
  accuracy of \emca\ calculations to about 0.2~meV.

\begin{table}
\caption{Convergence of \emca\ obtained via the \kkr\ code with the
  angular momentum cutoff
  $\ell_{\text{max}}^{\text{(KKR)}}$. \emca\ was evaluated by
  subtracting total energies. \label{lmax}}
\begin{ruledtabular}
\begin{tabular}{cc}
$\ell_{\text{max}}^{\text{(KKR)}}$ &$E_{\text{MCA}}$ (meV)\\
\hline
 2  &  1.289  \\
 3  &  3.101  \\
 4  &  3.437  \\
 5  &  3.407  \\
 6  &  3.217  \\
 7  &  3.039  \\
\end{tabular}
\end{ruledtabular}
\end{table}

%%---%%---%%---%%---%%---%%---%%---%%---%%---%%---%%---%%---%%---%%

\subsection{Convergence of \wien\ calculations with
  $R_{\text{MT}}K_{\text{max}}$ } 

\label{sec-rk}

An important parameter for the FLAPW calculations is the size of the
basis set.  It can be controlled by the $R_{\text{MT}}K_{\text{max}}$
product. The value $R_{\text{MT}}K_{\text{max}}$ = 7.0 is set by
default in {\sc wien}2k.  We increased the product
$R_{\text{MT}}K_{\text{max}}$ step by step from 6.0 up to 11.0 and
calculated the MCA energy.  The results are shown in Tab.~\ref{RKmax}.
It is clear from this that reliable values for the MCA energy can be
obtained for a basis set determined by the
$R_{\text{MT}}K_{\text{max}}$=8.0 condition.

\begin{table}[h]
\caption{ Convergence of \emca\ obtained via the \wien\ code with
  $R_{\text{MT}}K_{\text{max}}$.  \emca\ was evaluated via subtracting
  total energies (second column) and via the magnetic force
  theorem (third column).  \label{RKmax} }
\begin{ruledtabular}
\begin{tabular}{ccc}
$R_{\text{MT}}K_{\text{max}}$ & $E_{\text{MCA}}$ (meV) & $E_{\text{MCA}}$ (meV)\\
  &  \multicolumn{1}{c}{via $E_{\text{tot}}$} &
   \multicolumn{1}{c}{via force th.}  \\
\hline
 6.0  &  2.851   &   2.772  \\
 7.0  &  3.046   &   2.954  \\
 8.0  &  3.051   &  2.967   \\
 9.0  &  3.081   &  2.900   \\
10.0  &  2.993   & 2.908   \\
11.0  &  3.013   & 2.917   \\
\end{tabular}
\end{ruledtabular}
\end{table}

%%---%%---%%---%%---%%---%%---%%---%%---%%---%%---%%---%%---%%---%%

\subsection{Stability of \wien\ calculations with respect to
  $R_{\text{MT}}$ variations} 
 
\label{sec-rmt}

Recently the stability of the results with respect to varying the
muffin-tin radii was adopted as an informative test whether the FLAPW
basis set is sufficient or not.  Namely, in this way one changes the
regions where the wave functions are expanded in terms of plane waves
and where they are expanded in terms of atomic-like functions.  Only
if both expansions are appropriate the result will be stable against
this variation.  We adopted this test in our study, the results are
summarized in Tab.~\ref{RMT}.  We can see from a good agreement
between the MCA energies obtained for different muffin-tin radii
settings that the basis we used for our \wien\ calculations is
appropriate.

\begin{table}
\caption{ Dependence of \emca\ obtained via the \wien\ code on
  muffin-tin radii $R_{\text{MT}}$. \emca\ was evaluated via subtracting
  total energies (third column) and via the magnetic force
  theorem (fourth column). \label{RMT} }
\begin{ruledtabular}
\begin{tabular}{cccc}
$R_{\text{MT}}^{(\text{Fe})}$ (a.u.) & $R_{\text{MT}}^{(\text{Pt})}$
  (a.u.) &  $E_{\text{MCA}}$ (meV) &  $E_{\text{MCA}}$ (meV) \\
  &  & \multicolumn{1}{c}{via $E_{\text{tot}}$} &
   \multicolumn{1}{c}{via force th.}  \\
\hline
2.100 & 2.200 & 3.012 & 2.910  \\
2.180 & 2.280 & 3.083 & 2.973  \\
2.200 & 2.300 & 3.051 & 2.967  \\
2.220 & 2.320 & 3.004 & 2.848  \\
2.300 & 2.400 & 3.021 & 2.944  \\
\end{tabular}
\end{ruledtabular}
\end{table}

%%---%%---%%---%%---%%---%%---%%---%%---%%---%%---%%---%%---%%---%%

\subsection{Convergence of \kkr\ and \wien\ calculations with the
  number of $\bm{k}$-points} 
\label{sec-kpts}

A very important parameter is the number of $\bm{k}$-points used in
evaluating the integrals in the reciprocal space.  We performed
corresponding tests for both codes.  The dependence of \emca\ on the
number of $\bm{k}$-points in the full BZ is shown in
Tab.~\ref{kpoints}.  One can see that using about
100000~$\bm{k}$-points in the full Brillouin zone is sufficient to get
stable and reliable results.

\begin{table}
\caption{ Convergence of \emca\ calculated by the \kkr\ code (second
  column) and the 
  \wien\ code (third and fourth 
  columns)  with the number of $\bm{k}$-points in the full BZ.
 \emca\  (in meV) was evaluated via subtracting
  total energies (second and third columns) and via the magnetic force
  theorem (fourth column).  
   \label{kpoints}  }
\begin{ruledtabular}
\begin{tabular}{rccc} 
  &  \multicolumn{1}{c}{\emca [\kkr]}  & 
 \multicolumn{2}{c}{\emca [\wien]}  \\
\multicolumn{1}{c}{no.~of $\bm{k}$-points} &  
     \multicolumn{1}{c}{via $E_{\text{tot}}$} &
     \multicolumn{1}{c}{via $E_{\text{tot}}$} &
     \multicolumn{1}{c}{via force th.}  \\
\hline
  1000   &   2.894  &  2.996  &  2.897   \\
 10000   &   3.174  &  3.052  &  2.967   \\
 60000   &   3.129  &  3.009  &   2.896  \\
100000   &   3.101  &  3.051  &   2.967  \\
140000   &   3.091  &  3.024  &   2.966  \\
180000   &   3.092  &  2.944  &   3.008  \\
220000   &   3.099  &  3.090  &   2.848  \\
260000   &   3.103  &  3.001  &  2.897   \\
500000   &   3.099  &  2.997  &  2.894   \\
800000   &   3.096  &  2.989  &  2.848   \\ 
\end{tabular}
\end{ruledtabular}
\end{table}

%%---%%---%%---%%---%%---%%---%%---%%---%%---%%---%%---%%---%%---%%

\subsection{Convergence of \wien\ calculations with $E_{\text{max}}$}

\label{sec-emax}

When including the SOC within the second variation step, the size of
the new basis set is determined by the $E_{\text{max}}$ parameter
(Sec.~\ref{sec-relat}).  If $E_{\text{max}}$ is sufficiently
large, all scalar-relativistic eigenstates are involved.  The effect
of varying $E_{\text{max}}$ on the MCA energy in shown in
Tab.~\ref{Emax}.  One can see that if \emca\ is evaluated by
  means of the magnetic force theorem, it converges more quickly with
  $E_{\text{max}}$ than if \emca\ is evaluated via subtracting the
  total energies.  In both cases, nevertheless, the convergence is
  quite good. 

\begin{table}[b]
\caption{ Convergence of \emca\  
  obtained via the \wien\ code with $E_{\text{max}}$.  \emca\ was
  evaluated either by subtracting total energies (the second column)
  or by means of the magnetic force theorem (the third
  column).  \label{Emax} }
\begin{ruledtabular}
\begin{tabular}{rldd}
 \multicolumn{2}{c}{$E_{\text{max}}$ (Ry)} & 
\multicolumn{1}{c}{\emca\ (meV)}  & \multicolumn{1}{c}{\emca\ (meV)} \\
  & & \multicolumn{1}{c}{via $E_{\text{tot}}$} &
   \multicolumn{1}{c}{via force th.}  \\
\hline
   2  &                 & 3.117  & 2.955  \\
   5  &                 & 3.071  & 2.961  \\
  10  &                 & 3.064  & 2.965  \\
 100  & (all states)    & 3.051  & 2.967  \\
\end{tabular}
\end{ruledtabular}
\end{table}

%%---%%---%%---%%---%%---%%---%%---%%---%%---%%---%%---%%---%%---%%

% Produces the bibliography via BibTeX.

%%%%\bibliography{liter_co-on-pd}
%%% in our case:
\bibliography{biblo-fept} 

% File *.bbl inserted manually in order to avoid need for BibTeX
% cooperation;  this is an aid to PR publishing

\end{document}